\documentclass[letterpaper, 10pt, conference]{ieeeconf}      
\IEEEoverridecommandlockouts
\def\BibTeX{{\rm B\kern-.05em{\sc i\kern-.025em b}\kern-.08em
    T\kern-.1667em\lower.7ex\hbox{E}\kern-.125emX}}

\usepackage{cite,color,url}
\usepackage{amsmath,amssymb,amsfonts}
\usepackage{graphicx}
\usepackage{textcomp}

\usepackage{amsthm}

\let\labelindent\relax
\usepackage{enumitem}

\usepackage{epstopdf}

\usepackage{bm}
\usepackage{soul}
\usepackage{comment}

\newtheorem{theorem}{Theorem}

\newtheorem{lemma}{Lemma}
\newtheorem{assumption}{Assumption}

\newtheorem{problem}{Problem}
\newtheorem{remark}{Remark}

\allowdisplaybreaks

\begin{document}
\title{Probabilistic Performance Bounds for Randomized Sensor Selection in Kalman Filtering}
\author{Christopher I. Calle \qquad  Shaunak D. Bopardikar 
\thanks{The authors are with the Department of Electrical and Computer Engineering at the Michigan State University (MSU), East Lansing, MI, USA. Emails: \texttt{callechr@msu.edu, shaunak@egr.msu.edu}.
\newline
\indent This work was supported in part by NSF grant \# ECCS-2030556, GAANN grant \# P200A180025, the National GEM Fellowship, and MSU's University Enrichment Fellowship (UEF).}
}

\maketitle

\begin{abstract}
We consider the problem of randomly choosing the sensors of a linear time-invariant dynamical system subject to process and measurement noise. We sample the sensors independently and from the same distribution. We measure the performance of a Kalman filter by its estimation error covariance. Using tools from random matrix theory, we derive probabilistic bounds on the estimation error covariance in the semi-definite sense. We indirectly improve the performance of our Kalman filter for the maximum eigenvalue metric and show that under certain conditions the optimal sampling distribution that minimizes the maximum eigenvalue of the upper bound is the solution to an appropriately defined convex optimization problem. Our numerical results show the efficacy of the optimal sampling scheme in improving Kalman filter performance relative to the trivial uniform sampling distribution and a greedy sampling \textit{with replacement} algorithm.
\end{abstract}

\begin{keywords}
Sensor selection, Kalman filtering, Random matrix theory
\end{keywords}

\section{Introduction}
Recent years have seen rapid progress in technologies leading to the development of aerial, ground and underwater sensing platforms for a myriad of applications of significant societal impact. Examples include mobile platforms to study the development of severe weather \cite{doppler}, underwater sensors to detect and monitor the dispersal of chemical plumes \cite{chiny2007sensor}, and smart sensors to monitor traffic~\cite{zhang2015new}, to name a few. A common theme connecting these different scenarios is that the underlying quantity of interest evolves dynamically. Locations, concentrations and densities are the quantities of interest in our respective examples. In order to estimate the state of such quantities, we require sensors to observe our dynamical system. Such a need raises a practical question: \emph{Can one select the candidate sensors in an efficient manner and, simultaneously, provide provable guarantees on the estimation performance?} If the ubiquitous Kalman filter is the estimator for a linear time-invariant (LTI) state and measurement model subject to process and measurement noise, then this paper provides an affirmative answer. We assume the each sampled sensor is chosen \textit{with replacement} out of a sampling pool of candidate sensors. 




\subsection{Literature Review}



Sensor selection has a rich history in the control literature -- refer to \cite{muller1972} or \cite{van2001} for a survey of early works. In state estimation, metrics provide a quantitative approach to measuring the quality of a sensor selection. Recent works, such as \cite{summers2015submodularity} and \cite{pequito2016minimum}, discuss several metrics and efficient algorithms. In \cite{jawaid2015submodularity}, \cite{tzoumas2016sensor}, and \cite{zhang2017sensor}, the submodularity property of Kalman filter metrics are addressed.

In terms of sensor sampling techniques, the use of randomized sampling offers computational efficiency at the expense of yielding only probabilistic guarantees on the quality of the sampled sensor selection. Notable early works that studied random sampling of sensors and its effect on the estimation error covariance of the Kalman filter include \cite{gupta2006stochastic} and \cite{mo2011sensor}. More recent works include \cite{hashemi2018randomized} that analyze randomized greedy strategies and \cite{maske2018sensor} that perform sensor placement for nonlinear dynamics through a high-dimensional linear mapping of the feature space. 



Prior works \cite{mousavi2019measurable}, \cite{bopardikar2019randomized}, and \cite{bopardikar2021randomized} employed randomized sampling strategies to establish bounds on observability Gramian metrics. In \cite{bopardikar2019randomized} and \cite{bopardikar2021randomized}, matrix-valued concentration inequalities, like the Ahlswede-Winter inequality \cite{qiu2014sums}, are used to study dynamical systems with no process or measurement noise.
This paper investigates the more practical estimation problem, where the sensor network and the process it is attempting to estimate are corrupted by Gaussian noise. 
For this reason, we focus on the steady-state estimation error covariance of the Kalman filter to gauge state estimation performance.




\subsection{Contributions}


The contributions of this work are three-fold. First, using tools from random matrix theory, such as the Ahlswede-Winter inequality \cite{qiu2014sums}, we derive upper and lower bounds on the estimation error covariance with high probability. 
To our knowledge, our result is the first concentration inequality to bound the estimation error covariance in the semi-definite sense for an arbitrary sampling \textit{with replacement} policy.
Probabilistic guarantees in the semi-definite sense are appealing since they imply assurances on several metrics of significance in state estimation, such as the maximum eigenvalue, condition number, or trace of the estimation error covariance, to name a few. 

Second, under certain conditions, the optimal sampling distribution that minimizes the maximum eigenvalue of the upper bound is shown to be the solution to a convex optimization problem. We confirm that the optimal sampling distribution indirectly minimizes our actual metric of interest, the maximum eigenvalue of the steady-state estimation error covariance.

Third, we compare the state estimation performance of our optimal sampling distribution against the uniform sampling distribution and a greedy sampling \textit{with replacement} algorithm. 

\subsection{Outline of the paper}
This paper is organized as follows. We outline the linear dynamical system and sampling \textit{with replacement} scheme under consideration in Section~\ref{section:problem_formulation}. In Section~\ref{section:main_results}, we address the sensor selection problem in state estimation for the discrete-time Kalman filter. We derive probabilistic bounds on the steady-state error covariance and propose an optimal sampling distribution that indirectly improves
the state estimation performance for the maximum eigenvalue metric.
In Section~\ref{section:simulation_results}, we present numerical studies. Finally, we summarize our findings and identify directions for future research in Section~\ref{section:conclusion}. 
The proofs of all mathematical claims are presented in the appendix.


\section{Problem Formulation}
\label{section:problem_formulation}

\subsection{Notation}
We summarize the notation employed in this paper. Let~$\underline{\lambda}(\cdot)$ and $\overline{\lambda}(\cdot)$ denote the minimum and maximum eigenvalue of a Hermitian matrix argument, respectively. Let $I_n \in \mathbb{R}^{n \times n}$ denote the identity matrix and $\Delta^{n}$ denote the probability simplex in $\mathbb{R}^n$.

\subsection{Sampling Scheme}
\label{subsection:sampling_scheme}
For simplicity in notation, we assume each candidate sensor only outputs one measurement $y_{j,t} \in \mathbb{R}$. If a candidate sensor is modeled by a linear time-invariant measurement model corrupted by zero-mean Gaussian noise, i.e.,
\begin{align*}
y_{j,t} = \bm{c}_j^T x_t + \mu_{j,t},
\end{align*}
then the pair $(\bm{c}_j,\bm{\sigma}_j^2)$ is sufficient in completely describing the measurement properties of the $j$-th candidate sensor, where $\bm{c}_j^T$ is the sequence of weights that linearly relates the state $x_t$ to the output $y_{j,t}$ and $\bm{\sigma}_j^2$ is the measurement variance of Gaussian noise $\mu_{j,t}$. Let $\mathcal{X} := \{ (\bm{c}_{1},\bm{\sigma}_{1}^2) , \ldots , (\bm{c}_{n_c},\bm{\sigma}_{n_c}^2) \}$ denote the set of candidate sensor pairs and $n_c$ specify the number of candidate sensors under consideration. 

In our sensor selection problem, $n_s$ sensor pairs are chosen \textit{with replacement} from distribution $\mathcal{X}$. A sampling probability is assigned to each candidate sensor and the list of sampling probabilities is specified by the sampling distribution $p \in \Delta^{n_c}$. Let $\mathcal{S}  \in \{ 1 , \ldots , n_c \}^{n_s}$ denote the indices of the $n_s$ sampled candidate sensors. Throughout this paper, quantities that are either directly or indirectly dependent on our randomly generated sensor selection are accompanied by a subscript $\mathcal{S}$ notation. We assume the measurement properties $(\bm{c},\bm{\sigma}^2)$ of each candidate sensor are known prior to sampling.



\subsection{Model}
\label{subsection:problem_formulation_model}
Consider the tuple $(A,C_{\mathcal{S}},Q,R_{\mathcal{S}},I_m)$, an LTI state and measurement model subject to Gaussian noise, i.e.,
\begin{align}
\label{eqn:system_model_lti}
\begin{split}
x_{t+1} &= A \, x_t + w_t,   \\
y_t &= C_{\mathcal{S}} \, x_t + I_m \, v_t,
\end{split}
\end{align}
where $x_t \in \mathbb{R}^{m}$ is the state vector, $y_t \in \mathbb{R}^{n_s}$ is the output vector, and $n_s$ specifies the total number of observable measurements. Let $A \in \mathbb{R}^{m \times m}$ and $C_{\mathcal{S}} \in \mathbb{R}^{n_s \times m}$ denote the state and output matrix, respectively. Each row of $C_{\mathcal{S}}$ consists of row vector $c_{j}^{T}$, where $c_{j} \in \mathbb{R}^{m}$ relates the state $x_t$ to the output $y_{j,t}$ for the $j$-th sampled sensor. Let $w_t \sim \mathcal{N}(0,Q)$ and $v_t \sim \mathcal{N}(0,R_{\mathcal{S}})$ denote the process and measurement noise, respectively. Assume $\{ w_t \}_{t=0}^{\infty}$ and $\{ v_t \}_{t=0}^{\infty}$ are uncorrelated, zero-mean, white Gaussian processes. Additional assumptions on the noise properties of $w_t$ and $v_t$ are necessary for subsequent derivations.

\begin{assumption}
\label{assumption:covariance_Q,R}
Noise covariance matrices $Q$ and $R_{\mathcal{S}}$ are time-invariant and positive definite.
\end{assumption}

\begin{assumption}
\label{assumption:covariance_R}
Measurement covariance $R_{\mathcal{S}}$ is diagonal, i.e., $v_t$ consists of $n_s$ uncorrelated random variables, where $v_{j,t}$ and $\sigma_{j}^{2}$ denote the measurement noise and variance, respectively, corresponding to the $j$-th sampled sensor.
\end{assumption}

Note that our measurement model in \eqref{eqn:system_model_lti} is defined by the $n_s$ randomly chosen sensor pairs $(c_1,\sigma_1^{2}) , \dots , (c_{n_s},\sigma_{n_s}^{2})$ outlined in Section~\ref{subsection:sampling_scheme}.





\subsection{Sensor Selection for Kalman Filtering}
Under the assumptions of model linearity and Gaussian noise, the Kalman filter is a minimum mean squared error (MMSE) estimator that computes an optimal estimate of state $x_t$ in the mean-squared sense. If our measurement vector $y_t$ is available at each time instant $t$ for sensor fusion in a centralized manner, then the covariance \textit{information form} of the Kalman filter can be formulated into the following recursive equation,
\begin{align}
\label{eqn:filtered_covariance_recursion}
P_{\mathcal{S},t}^{-1} &= ( A P_{\mathcal{S},t-1}^{} A^{T} + Q )^{-1} + C_{\mathcal{S}}^{T} R_{\mathcal{S}}^{-1} C_{\mathcal{S}},
\end{align}
where $P_{\mathcal{S},t}$ denotes the filtered covariance of the state estimation error at time instant $t$. If $(A,C_{\mathcal{S}})$ and $(A,Q^{1/2})$ are detectable and stabilizable, respectively, then filtered error covariance $P_{\mathcal{S},t}$ converges to a steady-state solution~$P_\mathcal{S}$. Note the dependence of $C_\mathcal{S}^{T} R_\mathcal{S}^{-1} C_\mathcal{S}^{}$ on the row vector $c^T$ and measurement variance $\sigma^2$ of each randomly sampled sensor. In order to identify the contribution of each randomly sampled sensor in $C_\mathcal{S}^{T} R_\mathcal{S}^{-1} C_\mathcal{S}^{}$, Assumption~\ref{assumption:covariance_R} is established. If a symmetric, positive semi-definite, \textit{random} matrix $Z_j$ is generated by the pair $(c_{j},\sigma_{j}^{2})$ of the $j$-th randomly sampled sensor, i.e.,
\begin{align*}
Z_j = ( \sigma_{j}^{-1} c_{j} ) ( \sigma_{j}^{-1} c_{j} )^{T},
\end{align*}
then, under Assumption~\ref{assumption:covariance_R}, $C_\mathcal{S}^{T} R_\mathcal{S}^{-1} C_\mathcal{S}$ can be decomposed into a finite sum of independent and identically distributed (i.i.d.) random matrices,
\begin{align*}
C_\mathcal{S}^{T} R_\mathcal{S}^{-1} C_\mathcal{S}^{} = \sum_{j=1}^{n_s} c_{j}^{} \, \sigma_{j}^{-2} c_{j}^{T}
= \sum_{j=1}^{n_s} Z_{j}.
\end{align*}
Let $Z_1, \dots, Z_{n_s}$ denote independent copies of random variable $Z$, i.e., independently sampled matrices with the same distribution as $Z$. The expectation of random matrices $Z$ and $C_\mathcal{S}^{T} R_\mathcal{S}^{-1} C_\mathcal{S}^{}$ are given by
\begin{align*}
\mathbb{E}[Z] = \sum_{j=1}^{n_c} p_j \mathcal{Z}_{j} , \ \mathbb{E}[ C_\mathcal{S}^{T} R_\mathcal{S}^{-1} C_\mathcal{S}^{} ] = n_s \mathbb{E}[Z].
\end{align*}
Since the pairs $(c_1,\sigma_1^{2}) , \dots , (c_{n_s},\sigma_{n_s}^{2})$ are sampled from distribution $\mathcal{X}$ \textit{with replacement}, the symmetric, positive semi-definite, \textit{deterministic} matrix $\mathcal{Z}_{j}$ is constructed by the measurement properties $(\bm{c}_j,\bm{\sigma}_j^2)$, i.e.,
\begin{align*}
\mathcal{Z}_j = ( \bm{\sigma}_{j}^{-1} \bm{c}_{j} ) ( \bm{\sigma}_{j}^{-1} \bm{c}_{j} )^{T}
\end{align*}
for all $j \in \{ 1 , \ldots , n_c \}$.




\subsection{Problem Statement}
Our focus is on the following complementary problems.
\begin{problem}
\label{problem:bounds}
Given an arbitrary sampling distribution $p$, determine the upper and lower bounds on the steady-state error covariance $P_{\mathcal{S}}$ in the semi-definite sense.
\end{problem}

\begin{problem}
\label{problem:opt_sampling_dist}
Find an optimal sampling distribution $p^{*}$ that minimizes the maximum eigenvalue of the upper bound $P_U$ on the steady-state error covariance $P_{\mathcal{S}}$.
\end{problem}

Problem~\ref{problem:bounds} asks whether some minimal performance can be guaranteed with high probability, regardless of the sampling distribution under consideration. If such assurances exist, then the next question is whether there exists some ideal sampling scheme that optimizes our state estimation performance. Problem~\ref{problem:opt_sampling_dist} addresses the latter and asks how one can strategically choose a sensor selection to minimize a performance measure, specifically, the maximum eigenvalue of $P_{\mathcal{S}}$. Since $\overline{\lambda}( P_{\mathcal{S}} )$ is a random variable, it cannot be directly minimized. Instead, the maximum eigenvalue of upper bound $P_U$ is minimized in order to indirectly influence our actual metric of interest. Our solution to Problem~\ref{problem:bounds} and Problem~\ref{problem:opt_sampling_dist} are found in Section~\ref{subsection:steady_state_solution_guarantees} and Section~\ref{subsection:opt_sampling_dist}, respectively.

\section{Main Results}
\label{section:main_results}

First, we derive probabilistic bounds on the steady-state error covariance $P_{\mathcal{S}}$ in the semi-definite sense. Next, the expected steady-state solution $\mathbb{E}[P_{\mathcal{S}}]$ and its relation to the bounds $P_U$ and $P_L$ are explored.
Lastly, the sampling distribution $p^{*}$ that optimally minimizes $\overline{\lambda}( P_U )$ is obtained in order to indirectly minimize $\overline{\lambda}( P_{\mathcal{S}} )$ and improve our state estimation performance.

\subsection{Steady-State Solution Guarantees}
\label{subsection:steady_state_solution_guarantees}

Before establishing bounds on the steady-state solution $P_{\mathcal{S}}$, the filtered error covariance in the deterministic setting is investigated. If we assume $n_s$ sensors are chosen beforehand and not randomly sampled, then the output matrix and measurement covariance of an LTI system are deterministic. Lemma \ref{lemma:arbitrary_model_Y1_Y2_Y3} outlines the conditions required to deterministically upper and lower bound the filtered error covariance of an arbitrary LTI system in the semi-definite sense.

\begin{lemma}
\label{lemma:arbitrary_model_Y1_Y2_Y3}
\textbf{(Deterministic Bounds)}
Consider the following LTI systems, $(A,Y_3^{1/2},Q,\Pi_3,\Pi_3^{-1/2})$, $(A,\Gamma_2,Q,\Pi_2,I_m)$, and $(A,Y_1^{1/2},Q,\Pi_1,\Pi_1^{-1/2})$, and define their filtered error covariance matrices, i.e.,
\begin{align*}
{P}_{1,t}^{-1} &= (A {P}_{1,t-1}^{} A^{T} + Q)^{-1} + Y_{3},   \\
{P}_{2,t}^{-1} &= (A {P}_{2,t-1}^{} A^{T} + Q)^{-1} + Y_{2},   \\
{P}_{3,t}^{-1} &= (A {P}_{3,t-1}^{} A^{T} + Q)^{-1} + Y_{1},
\end{align*}
respectively, such that $Y_{i} = \Gamma_{i}^{T} \Pi_{i}^{-1} \Gamma_{i}^{}$ for all $i \in \{1,2,3\}$. If the following conditions are satisfied,
\begin{enumerate}[ label=(\roman*) , leftmargin=2.5mm , align=left ]
\item [(C1)] $0 \preceq Y_{1} \preceq Y_{2} \preceq Y_{3}$, and
\item [(C2)] $0 \preceq {P}_{1,-1} \preceq {P}_{2,-1} \preceq {P}_{3,-1}$, and
\item [(C3)] $(A, Y_{3}^{1/2} )$, $(A, \Gamma_{2} )$, and $(A, Y_{1}^{1/2} )$ are detectable,
\end{enumerate}
then $P_{1} \preceq P_{2} \preceq P_{3}$ and
\begin{align*}
P_{1,t-1} \preceq P_{2,t-1} \preceq P_{3,t-1}, \ \forall t \geq 0.
\end{align*}
\end{lemma}

Lemma \ref{lemma:arbitrary_model_Y1_Y2_Y3} in tandem with the Ahlswede-Winter inequality, 
Theorem~\ref{theorem:Ahlswede_Winter} in the appendix, yields probabilistic bounds in the semi-definite sense on the steady-state solution $P_{\mathcal{S}}$ for an arbitrary sampling distribution $p$.

\begin{theorem}
\label{theorem:steady-state_bounds}
\textbf{(Probabilistic Steady-State Bounds)} \\
Let $m, n_s \in \mathbb{N}$, $\delta \in (0,1)$, $\rho \in \big{[} 1 , \frac{n_s \epsilon^2}{4 \log{ ( 2m / \delta ) }} \big{)}$ and
\begin{equation}
\label{eqn:epsilon}
\epsilon = \sqrt{ \frac{ 4 \rho }{ n_s } \log{ \frac{2m}{\delta} } } \in (0,1)
\end{equation}
for specified sampling distribution $p$, such that $\mathcal{Z}_j \preceq \rho \, \mathbb{E}[Z]$ for all $j \in \{ 1 , \ldots , n_c \}$. Assume $(A, \mathbb{E}[Z]^{1/2})$ and $(A,C_{\mathcal{S}})$ are detectable and $(A,Q^{1/2})$ is stabilizable. If $P_U$ and $P_L$ denote steady-state solutions, i.e.,
\begin{align*}
P_{U}^{-1} &= (A P_{U}^{} A^{T} + Q)^{-1} + (1-\epsilon) n_s \mathbb{E}[Z],   \\
P_{L}^{-1} &= (A P_{L}^{} A^{T} + Q)^{-1} + (1+\epsilon) n_s \mathbb{E}[Z],
\end{align*}
and
\begin{align*}
P_{\mathcal{S}}^{-1} = (A P_{\mathcal{S}}^{} A^{T} + Q)^{-1} + C_{\mathcal{S}}^{T} R_{\mathcal{S}}^{-1} C_{\mathcal{S}}^{},
\end{align*}
then the steady-state error covariance $P_{\mathcal{S}}$ satisfies
\begin{align*}
\mathbb{P} [ P_{L} \preceq P_{\mathcal{S}} \preceq P_{U} ] \geq (1 - \delta).
\end{align*}
\end{theorem}

A few comments are summarized below.

\begin{remark}
\label{remark:bounds_sensor_sampling_parameters}
If $m$ is fixed, then the only parameters that can be tuned to guarantee $\epsilon \in (0,1)$ are $\delta$, $n_s$, and $\rho$. Quantities $\delta$ and $n_s$ can be easily tuned since they are user-specified. In contrast, tuning the value of $\rho$ is a non-trivial problem due to its dependence on the sampling distribution $p$.
\end{remark}


\begin{remark}
\label{remark:detectability_assums}
The detectability assumptions of Theorem~\ref{theorem:steady-state_bounds} can be satisfied by one of the following sufficient conditions:
\begin{enumerate}[ label=(\roman*) , leftmargin=2.5mm , align=left ]
    \item [(S1)] $(A,\bm{c}_j)$ is observable for all $j \in \{ 1 , \ldots , n_c \}$, or
    \item [(S2)] Prior to randomly selecting $n_s$ sensors as outlined in Section \ref{section:problem_formulation}, if $n_a$ additional sensors are first strategically sampled from the sampling pool of candidate sensors and shown to collectively guarantee observability of the system, then the steady-state solutions $P_U$, $P_L$, and $P_{\mathcal{S}}$ will always exist, regardless of the sampling distribution $p$, assuming the stabilizability condition is satisfied.
\end{enumerate}
\end{remark}

\begin{remark}
From the dependence in \eqref{eqn:epsilon}, we conclude that for the analysis to be applicable, we require
\begin{align*}
n_s \geq \frac{4\rho}{\epsilon^2} \log\frac{2m}{\delta}.    
\end{align*}
Thus, the number of samples show a logarithmic dependence on $m$ and $1/\delta$, which is reasonable. However, the $1/\epsilon^2$ dependence is a consequence of sampling with replacement and the central limit theorem which is the key result used in the proof of Theorem~\ref{theorem:Ahlswede_Winter}, the Ahlswede-Winter inequality, that is employed to establish Theorem~\ref{theorem:steady-state_bounds}.
\end{remark}

In order to measure the average state estimation performance 
of our sampling scheme in Section~\ref{section:problem_formulation}, we introduce an analytical lower bound on the expectation of the steady-state solution $P_{\mathcal{S}}$ that closely approximates it.

\begin{lemma}
\label{lemma:lower_bound_L_&_E_P}
\textbf{(Analytical Lower Bound)}
If $(A, \mathbb{E}[Z]^{1/2})$ and $(A,Q^{1/2})$ are detectable and stabilizable, respectively, and if $L$ denotes the solution to the following,
\begin{align*}
L^{-1} = (A L A^{T} + Q)^{-1} + n_s \mathbb{E}[Z],
\end{align*}
then $L \preceq \mathbb{E}[ P_{\mathcal{S}} ]$.
\end{lemma}

Lemma~\ref{lemma:lower_bound_L_&_E_P} is used to explain how the bounds $P_U$ and $P_L$ of Theorem~\ref{theorem:steady-state_bounds} are related to the expected steady-state error covariance $\mathbb{E}[ P_{\mathcal{S}} ]$.

\begin{remark}
\label{remark:bounds_P_I}
In the limit as $\epsilon$ tends to $0$, $P_{U}$ and $P_{L}$ tend to identical solutions, i.e.,
\begin{align*}
\lim_{\epsilon \rightarrow 0} P_{U}^{-1} &= (A P_{U}^{} A^{T} + Q)^{-1} + n_s \mathbb{E}[Z],   \\
\lim_{\epsilon \rightarrow 0} P_{L}^{-1} &= (A P_{L}^{} A^{T} + Q)^{-1} + n_s \mathbb{E}[Z].
\end{align*}
For this special case, $P_{U}$ and $P_{L}$ are denoted by $P_{I}$, and
\begin{align}
\label{eqn:P_I_inv}
P_{I}^{-1} &= (A P_{I}^{} A^{T} + Q)^{-1} + n_s \mathbb{E}[Z].
\end{align}
A simple comparison of Lemma~\ref{lemma:lower_bound_L_&_E_P} and \eqref{eqn:P_I_inv} shows that $L$ and $P_{I}$ are identical. This implies that $P_{U}$ and $P_{L}$ bound a lower bound of $\mathbb{E}[ P_{\mathcal{S}} ]$, denoted as $L$, for all $\epsilon \in (0,1)$. As $\epsilon$ decreases (increases), $P_{U}$ and $P_{L}$ converge towards (diverge from) $L$ in the semi-definite sense.
\end{remark}

\subsection{Optimal Sampling Scheme}
\label{subsection:opt_sampling_dist}

In Section~\ref{subsection:steady_state_solution_guarantees}, the steady-state error covariance is bounded in the probabilistic sense for an arbitrary sampling distribution $p$. Due to its dependence on a randomly chosen sensor selection $\mathcal{S}$, the steady-state solution $P_{\mathcal{S}}$ cannot directly be influenced. Instead, the bounds $P_U$ and $P_L$ of Theorem 1 must be used to indirectly affect state estimation performance. In this section, the maximum eigenvalue of the steady-state solution $P_{\mathcal{S}}$ is the performance metric of interest. By minimizing the maximum eigenvalue of the upper bound $P_U$, $\overline{\lambda}( P_{\mathcal{S}} )$ is similarly minimized with high probability.


\begin{theorem}
\label{theorem:opt_sampling_dist}
\textbf{(Optimal Sampling Distribution)}
A sampling distribution $p_{\rho}^{*} = \{ p_{i}^{*} \}_{i=1}^{n_c}$ that optimally minimizes $\overline{\lambda}( P_{U} )$ with respect to a selected $\rho \in \big{[} 1 , \frac{n_s \epsilon^2}{4 \log{ ( 2m / \delta ) }} \big{)}$ and an arbitrarily small $\eta > 0$ is computed by solving the following semi-definite program (SDP).
\begin{equation*}
\label{eqn:opt_problem_01}
\begin{split}
\max_{ \lambda , X , \{ p_{i} \}_{i=1}^{n_c} } \quad & \lambda   \\
\mathrm{s.t.} \hspace{5mm} \quad 
& \lambda > 0 , \ X \succeq \eta I_m , \ \{ p_{i} \}_{i=1}^{n_c} \in \Delta^{n_c}   \\
& \epsilon = \sqrt{ \frac{ 4 \rho }{ n_s } \log{ \frac{2m}{\delta} } } \in (0,1)   \\
&
\begin{bmatrix}
X + A^{T} Q^{-1} A & ( A^{T} Q^{-1} )   \\
( A^{T} Q^{-1} )^{T} & \Gamma_{3}
\end{bmatrix}
\succeq 0   \\
& \Gamma_{3} = Q^{-1} + (1-\epsilon) n_s \sum_{j=1}^{n_c} p_{j} \mathcal{Z}_{j} - \lambda I_m  \\
& \mathcal{Z}_i \preceq \rho \, \sum_{j=1}^{n_c} p_{j} \mathcal{Z}_{j} , \ \forall i \in \{ 1 , \ldots , n_c \}
\end{split}
\end{equation*}
\end{theorem}

Theorem \ref{theorem:opt_sampling_dist} computes the sampling distribution $p_{\rho}^{*}$ for a selected $\rho$ and $\eta$. In order to find the sampling distribution $p^{*}$ that optimally minimizes $\overline{\lambda}( P_{U} )$, irrespective of $\rho$, a search algorithm is necessary. 

\begin{remark}
\label{remark:optimal_procedure}
One should expect that minimizing $\rho$ will minimize $\overline{\lambda}( P_{U} )$ upon inspection of constraint \eqref{eqn:epsilon}, since minimizing $\rho$ minimizes $\epsilon$ and, subsequently, tightens the bounds outlined in Theorem \ref{theorem:steady-state_bounds}. Though this heuristic can be used to identify a relatively minimal $\overline{\lambda}( P_{U} )$, it cannot be guaranteed to find the global minimum. Instead, the optimal $\rho^{*}$ that globally minimizes $\overline{\lambda}( P_{U} )$ can be found incrementally. By employing a binary or bisection search procedure throughout the feasible regime of $\rho$, Theorem \ref{theorem:opt_sampling_dist} can be consecutively applied to find the $\rho$ that globally minimizes $\overline{\lambda}( P_{U} )$ within a predefined constant~$\gamma$ of the optimal $\rho^{*}$.
\end{remark}

\section{Simulation Results}
\label{section:simulation_results}




In this section, the optimal sampling distribution computed in Section  \ref{subsection:opt_sampling_dist} is demonstrated to substantially minimize the maximum eigenvalue of upper bound $P_U$ relative to a trivial uniform sampling distribution. We also demonstrate that relative to a greedy sampling \textit{with replacement} algorithm the average state estimation performance of our optimal sampling distribution is consistently superior for the maximum eigenvalue metric.

In our numerical analysis, the state dimension $m = 3$, the number of candidate sensors $n_c = 200$, and $\delta = 0.10$. We assume the process covariance matrix $Q = 0.5 \, I_m$ and the measurement noise variance of each candidate sensor is identical, such that $\bm{\sigma}_{j}^{2} = 0.5$ for all $j \in \{ 1 , \ldots , n_c \}$. The entries of state matrix $A$ and output vector $\bm{c}$ for each candidate sensor are chosen independently and uniformly at random out of the interval $[0,1]$. Detectability conditions of Theorem~\ref{theorem:steady-state_bounds} are satisfied by verifying that the synthetically generated pair $(A,\bm{c}_{j})$ is observable for all $j \in \{ 1 , \ldots , n_c \}$.

\begin{figure}[h]
    \centering
    \includegraphics[width=75mm]{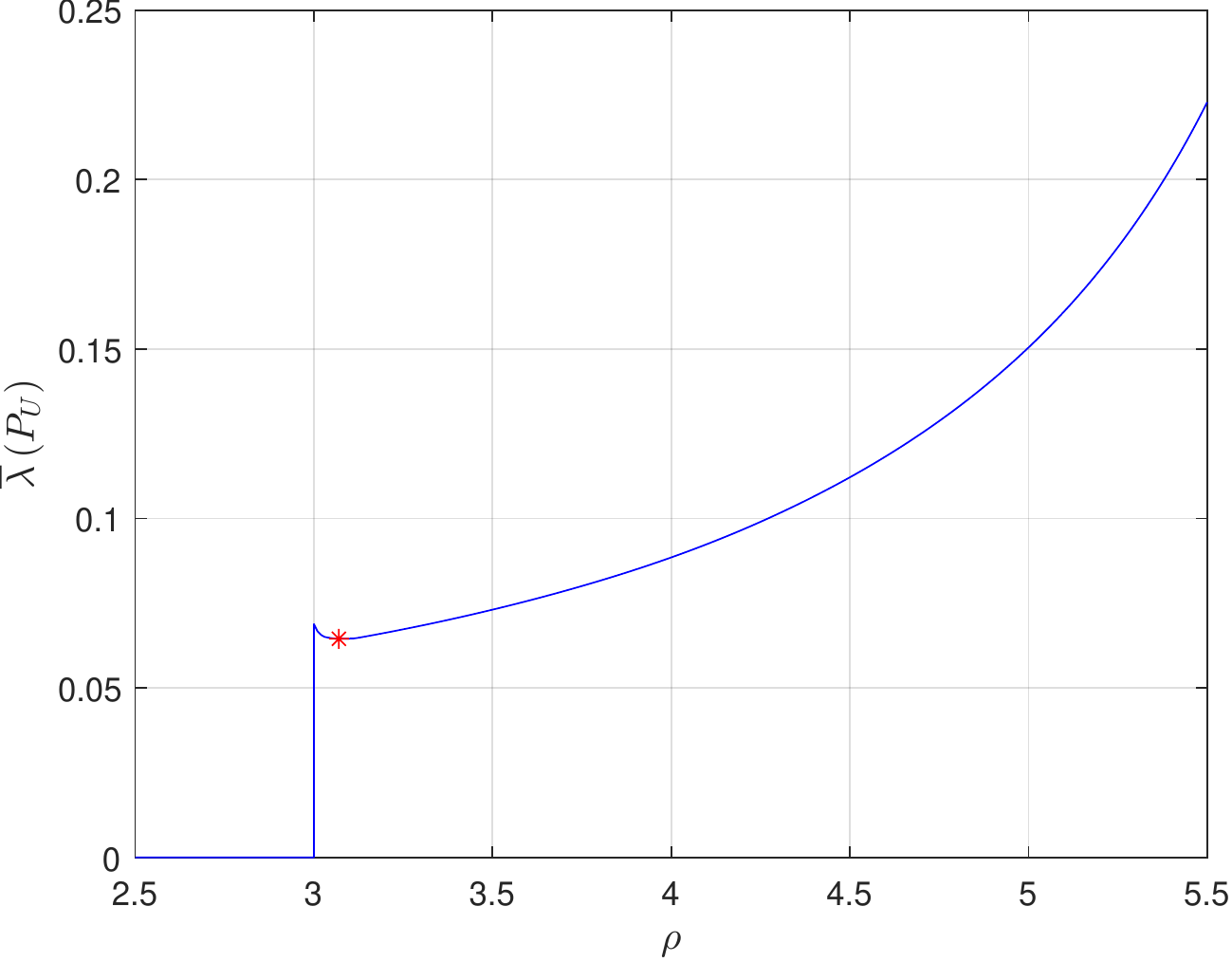}
    \caption{Maximum eigenvalue of the upper bound $P_U$ for a limited regime of $\rho$ values. Non-zero values of $\overline{\lambda}( P_{U} )$ indicate feasible $\rho$ values. The red asterisk locates the optimal $\rho^{*}$ value that globally minimizes $\overline{\lambda}( P_{U} )$.}
    \label{figure:rho_vs_lambda_max}
\end{figure}

In Figure~\ref{figure:rho_vs_lambda_max}, the optimal sampling distribution $p_{\rho}^{*}$ and its corresponding $\overline{\lambda}( P_{U} )$ are computed using Theorem~\ref{theorem:opt_sampling_dist} for varying $\rho$ values and a fixed number of sampled sensors $n_s = 100$. Figure~\ref{figure:rho_vs_lambda_max} confirms our discussion in Remark~\ref{remark:optimal_procedure}, such that minimizing $\rho$ tends to minimize $\overline{\lambda}( P_{U} )$ in general. Furthermore, the nature of the maximum eigenvalue curve over the regime of feasible $\rho$ values motivates us to conjecture that $\overline{\lambda}(P_U)$ is a convex function of $\rho$. If proven true, the search procedure outlined in Section \ref{subsection:opt_sampling_dist} would be obsolete and the globally minimum $\overline{\lambda}(P_U)$ could be solved directly with minor alterations to Theorem~\ref{theorem:opt_sampling_dist}. 

\begin{figure}[h]
    \centering
    \includegraphics[width=75mm]{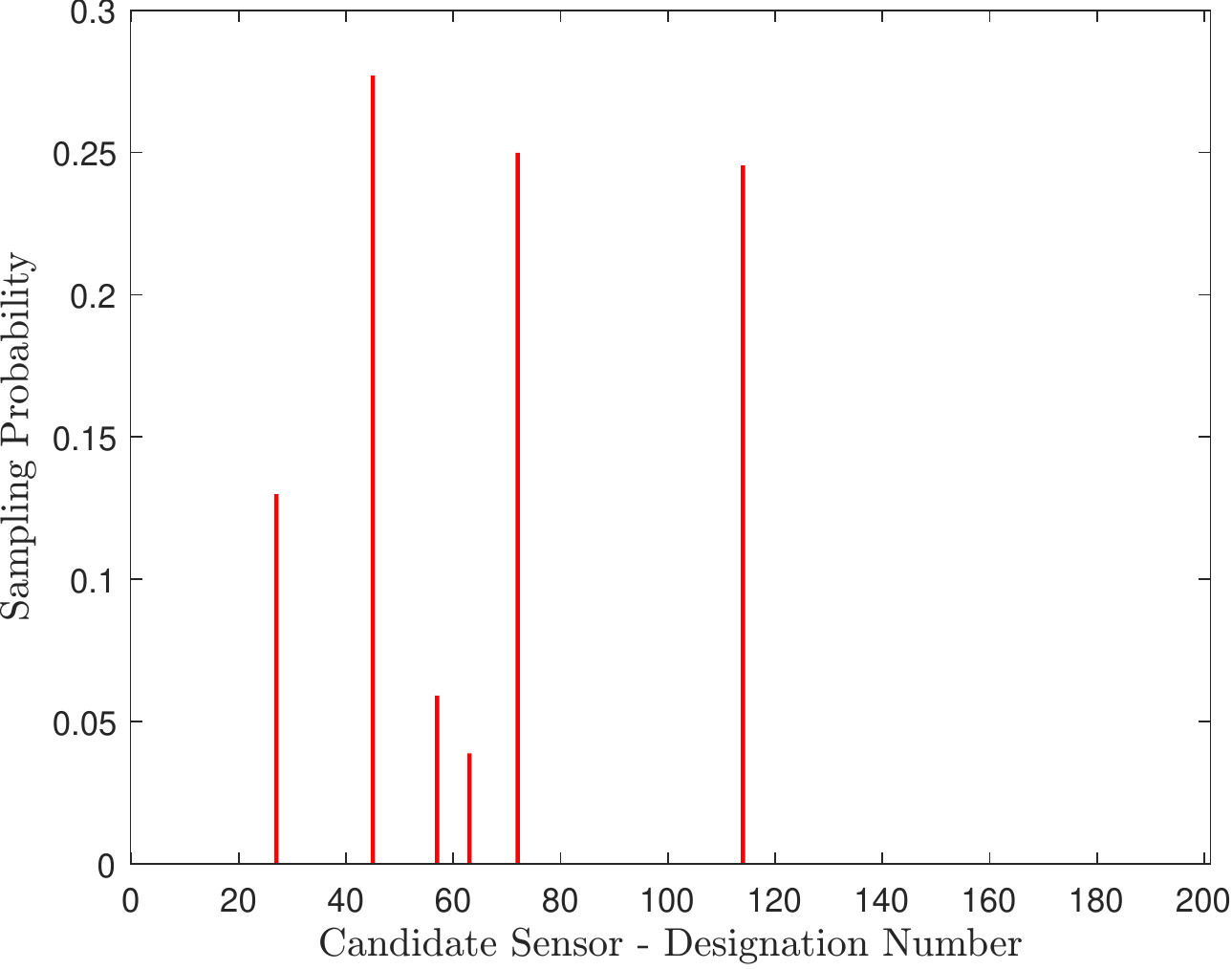}
    \caption{Optimal sampling distribution $p^{*}$ that globally minimizes $\overline{\lambda}( P_{U} )$.}
    \label{figure:opt_prob_dist}
\end{figure}

In Figure~\ref{figure:opt_prob_dist}, the optimal sampling distribution $p^{*}$ is plotted. Note that sampling distribution $p^{*}$ is sparse and Figure~\ref{figure:opt_prob_dist} identifies the small subset of candidate sensors collectively responsible for minimizing $\overline{\lambda}(P_U)$ by the greatest margin.

\begin{figure}[h]
    \centering
    \includegraphics[width=75mm]{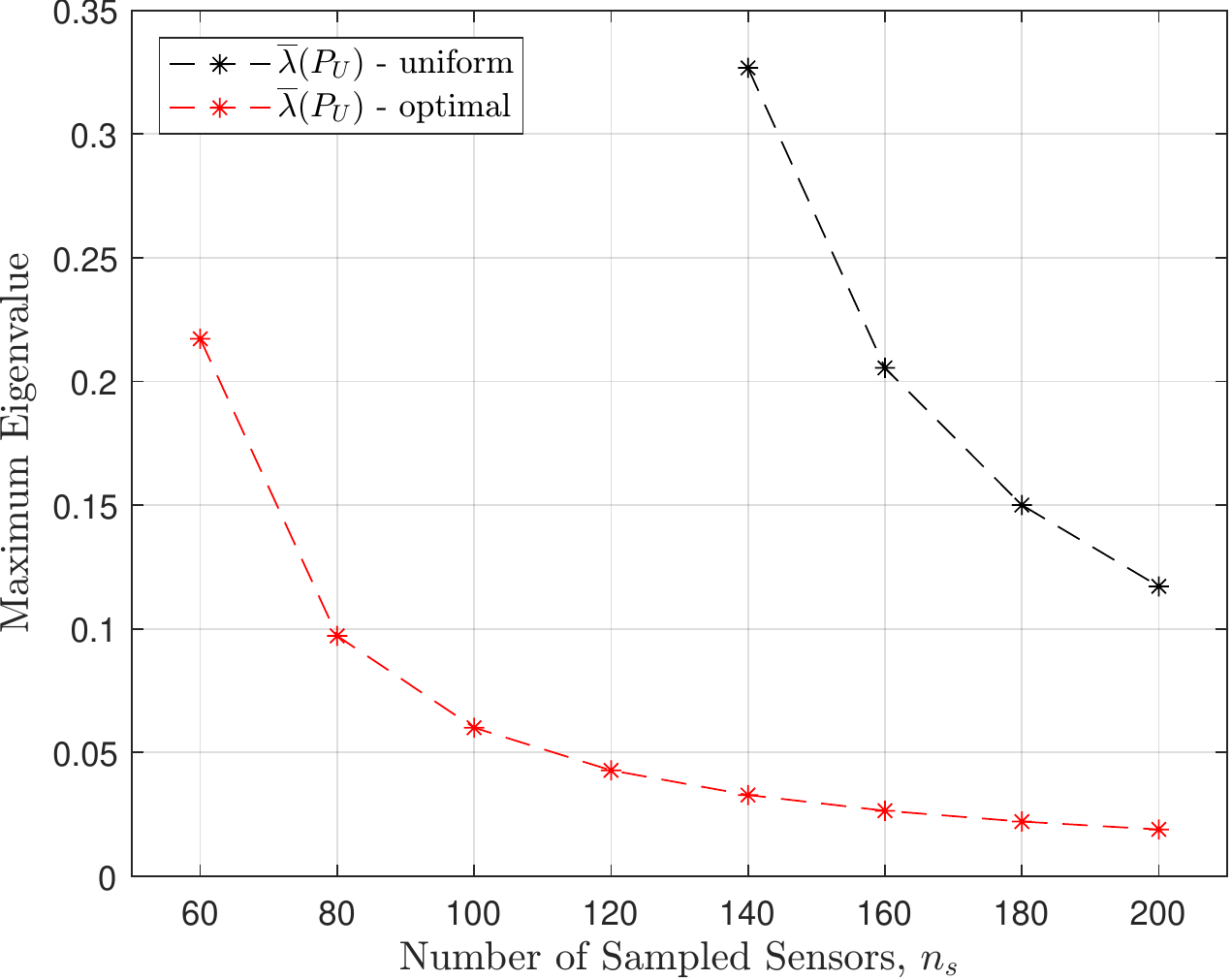}
    \caption{A numerical comparison between the uniform (black-star) and optimal (red-star) sampling distribution for the maximum eigenvalue of the upper bound $P_U$ for varying number of sampled sensors $n_s$.}
    \label{figure:ns_vs_lambda_max}
\end{figure}

In Figure~\ref{figure:ns_vs_lambda_max}, the sampling distribution $p^{*}$ that globally minimizes $\overline{\lambda}( P_{U} )$ is computed for varying number of sampled sensors and compared as a benchmark against the $\overline{\lambda}( P_{U} )$ curve for a trivial uniform sampling distribution. Figure~\ref{figure:ns_vs_lambda_max} shows that the uniform sampling distribution is only applicable for a limited regime of sampled sensors. In fact, if too few sensors are sampled, then the probabilistic guarantees of Theorem~\ref{theorem:steady-state_bounds} no longer hold. In contrast, the $\overline{\lambda}( P_{U} )$ curve for the optimal sampling distribution $p^{*}$ requires significantly fewer sampled sensors to substantially minimize the maximum eigenvalue of upper bound $P_{U}$.

\begin{figure}[h]
    \centering
    \includegraphics[width=75mm]{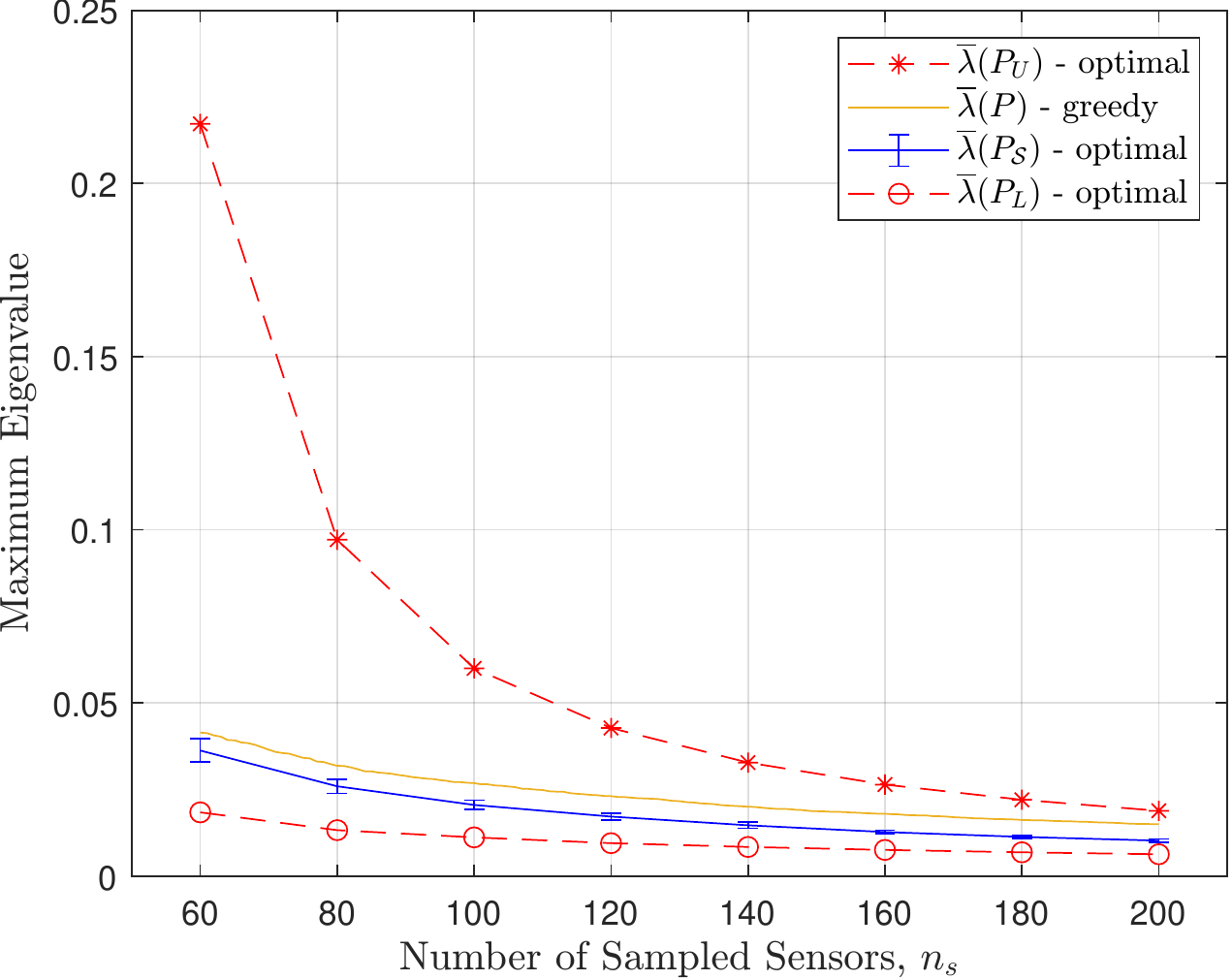}
    \caption{Maximum eigenvalue for varying number of sampled sensors $n_s$, such that the maximum eigenvalue of $P_U$ and $P_L$ are indicated by the (red-star) and (red-circle) curve, respectively. The maximum eigenvalue of a steady-state error covariance obtained per a greedy sampling \textit{with replacement} algorithm is indicated by the (yellow-line) curve. The average maximum eigenvalue of $P_{\mathcal{S}}$ is indicated by the (blue-line) curve and variability of $\overline{\lambda}( P_\mathcal{S} )$ is captured by the standard deviation. The error bars indicate $\pm$ one standard deviation.}
    \label{figure:greedy_comparison}
\end{figure}

In Figure~\ref{figure:greedy_comparison}, quantities $\overline{\lambda}( P_U )$, $\overline{\lambda}( P_L )$, and $\overline{\lambda}( P_\mathcal{S} )$ are plotted for varying number of sampled sensors and compared against the $\overline{\lambda}( P )$ obtained via a greedy sampling \textit{with replacement} scheme. For each $n_s$ the maximum eigenvalue of bounds $P_U$ and $P_L$ are computed using their corresponding optimal sampling distribution $p^{*}$. Similarly, for each $n_s$ and corresponding optimal distribution $p^{*}$, the average maximum eigenvalue of steady-state solution $P_{\mathcal{S}}$ is estimated by 100 Monte Carlo trials. In the greedy algorithm, $n_s$ sensors are sampled \textit{with replacement} out of the sampling pool of candidate sensors. At each sampling instant, the candidate sensor that minimizes $\overline{\lambda}( P )$ is greedily chosen. Figure~\ref{figure:greedy_comparison} shows that the average $\overline{\lambda}( P_\mathcal{S} )$ generated by sampling distribution $p^{*}$ is consistently smaller than the $\overline{\lambda}( P )$ of the greedy algorithm.

\section{Conclusion}
\label{section:conclusion}

In this paper, we consider the sensor selection problem under the context of state estimation for the discrete-time Kalman filter.
Novel bounds on the steady-state error covariance of a randomly sampled sensor selection were derived in the probabilistic sense using tools from random matrix theory.
We confirmed that the sampling distribution that minimizes the maximum eigenvalue of the upper bound indirectly minimizes the maximum eigenvalue of the steady-state error covariance. Our simulations demonstrated that the optimal sampling distribution significantly outperforms the trivial uniform sampling distribution in terms of the maximum eigenvalue of the upper bound. A numerical analysis showed that the maximum eigenvalue of the steady-state error covariance generated by the optimal sampling distribution consistently outperforms on-average a greedy sampling \textit{with replacement} algorithm. Our results are expected to be significant in the analysis of large sensor networks, since manually choosing the sampling distribution that minimizes a non-trivial objective function is infeasible.

Future directions include extending our analytical guarantees on state estimation performance to the constrained setting, where each candidate sensor is limited in availability and cannot be sampled \textit{with replacement} indefinitely.


\bibliographystyle{ieeetr} 
\bibliography{references}

\appendix 

This section contains the mathematical proofs of all the claims presented in this paper. A useful inequality from random matrix theory, known as the Ahlswede-Winter inequality, allows us to bound sums of independent positive semi-definite matrices.
\begin{theorem}
\label{theorem:Ahlswede_Winter}
\textbf{(Ahlswede-Winter Inequality)}
Let $Z$ be a random, symmetric, positive semi-definite $m \times m$ matrix. Define $U=\mathbb{E}[Z]$ and suppose that $Z\preceq \rho \, U$ almost surely, for some scalar $\rho \geq 1$. Let $Z_1, \dots, Z_{n_s}$ denote independent copies of $Z$, i.e., independently sampled matrices with the same distribution as $Z$. For any $\epsilon \in (0,1)$, we have
\begin{align*}
\mathbb{P} \Bigg{[} (1-\epsilon) \, U \preceq \frac{1}{n_s} \sum_{j=1}^{n_s} Z_j \preceq (1+\epsilon) \, U \Bigg{]} \geq \Big{(} 1 - 2m \, { \mathrm{e} }^{ -\frac{ \epsilon^2 n_s }{ 4 \rho } } \Big{)}.
\end{align*}
\end{theorem}




Theorem~\ref{theorem:Ahlswede_Winter} is Corollary 2.2.2 in \cite{qiu2014sums} with minor alterations to the notation.



\subsection{Proof of Lemma~\ref{lemma:arbitrary_model_Y1_Y2_Y3}}

Let the tuple $(A,\Gamma,Q,\Pi,I_m)$ denote a discrete-time LTI system, such that $\Gamma$ and $\Pi$ are \textit{deterministic} and not generated per a randomized sensor sampling scheme, as outlined in Section~\ref{subsection:sampling_scheme}. Let $w_t \sim \mathcal{N}(0,Q)$ and $v_t \sim \mathcal{N}(0,\Pi)$ denote the process and measurement noise, respectively, and define the filtered covariance matrix $P_{t}$ as the following,
\begin{align*}
P_{t}^{-1} = ( A P_{t-1}^{} A^T + Q )^{-1} + Y,
\end{align*}
such that $Y = \Gamma^T \Pi^{-1} \Gamma$.
If $(A,\Gamma)$ and $(A,Q^{1/2})$ are detectable and stabilizable, respectively, then the filtered covariance matrix converges to its steady-state solution $P$. It can be shown that the filtered covariance matrix of augmented discrete-time LTI system $(A,\Pi^{-1/2} \Gamma,Q,\Pi,\Pi^{-1/2})$, alternatively denoted as $(A,Y^{1/2},Q,\Pi,\Pi^{-1/2})$, i.e.,
\begin{align*}
\begin{split}
x_{t+1} &= A \, x_t + w_t,   \\
y_t &= \Pi^{-1/2} \Gamma \, x_t + \Pi^{-1/2} \, v_t,
\end{split}
\end{align*}
is identical to the filtered covariance matrix of tuple $(A,\Gamma,Q,\Pi,I_m)$. In contrast, the filtered covariance matrix corresponding to $(A,Y^{1/2},Q,\Pi,\Pi^{-1/2})$ converges to its steady-state solution if $(A,Y^{1/2})$ and $(A,Q^{1/2})$ are detectable and stabilizable, respectively.

In this lemma, we consider the following discrete-time LTI systems, $(A,Y_3^{1/2},Q,\Pi_3,\Pi_3^{-1/2})$, $(A,\Gamma_2,Q,\Pi_2,I_m)$, and $(A,Y_1^{1/2},Q,\Pi_1,\Pi_1^{-1/2})$, and define their filtered covariance matrices, i.e.,
\begin{align*}
{P}_{1,t}^{-1} &= (A {P}_{1,t-1}^{} A^{T} + Q)^{-1} + Y_{3},   \\
{P}_{2,t}^{-1} &= (A {P}_{2,t-1}^{} A^{T} + Q)^{-1} + Y_{2},   \\
{P}_{3,t}^{-1} &= (A {P}_{3,t-1}^{} A^{T} + Q)^{-1} + Y_{1},
\end{align*}
respectively, such that $Y_{i} = \Gamma_{i}^{T} \Pi_{i}^{-1} \Gamma_{i}^{}$ for all $i \in \{1,2,3\}$. Filtered covariance matrices $P_{1,t}$, $P_{2,t}$, and $P_{3,t}$ converge to their steady-state solution if $(A, {Y_3}^{1/2} )$, $(A, {\Gamma_2} )$ and $(A, {Y_1}^{1/2} )$ are detectable, respectively, and $(A,Q^{1/2})$ is stabilizable. Let $P_{1,-1}$, $P_{2,-1}$, and ${P}_{3,-1}$ denote the \textit{initial} estimate of $P_{1,t-1}$, $P_{2,t-1}$, and ${P}_{3,t-1}$, respectively, and assume the following initial condition,
\begin{align}
\label{eqn:P_2_-1_bounds}
0 \preceq P_{1,-1} \preceq P_{2,-1} \preceq P_{3,-1}.
\end{align}
Let us assume $Y_2$ satisfies
\begin{align}
\label{eqn:inequality_Y}
0 \preceq Y_{1} \preceq Y_{2} \preceq Y_{3}.
\end{align}
In order to prove the semi-definite bounds,
\begin{align}
\label{eqn:P_2_t-1_bounds}
P_{1,t-1} \preceq P_{2,t-1} \preceq P_{3,t-1},
\end{align}
hold for all $t \geq 0$, the following inequality, $P_{1,t-1} \preceq P_{2,t-1}$, is first shown to hold for all $t \geq 0$. Inequality \eqref{eqn:inequality_Y} and a fundamental property of Hermitian matrices \cite{bhatia2013matrix}, explicitly referred to as the Conjugation Rule in \cite{tropp2015introduction}, satisfies
\begin{align*}
A P_{1,t-1}^{} A^{T} \preceq A P_{2,t-1}^{} A^{T}.
\end{align*}
Inequality \eqref{eqn:inequality_Y} and Assumption~\ref{assumption:covariance_Q,R} guarantee that $Q~\succ~0$ and $Y_2~\preceq~Y_3$, respectively, and thereby satisfies
\begin{align*}
( A P_{1,t-1}^{} A^{T} + Q )^{-1} + Y_{3} \succeq ( A P_{2,t-1}^{} A^{T} + Q )^{-1} + Y_{2}
\end{align*}
and, consequently, $P_{1,t} \preceq P_{2,t}$. A similar derivation holds for inequality, $P_{2,t-1} \preceq P_{3,t-1}$, thereby, proving that inequality \eqref{eqn:P_2_t-1_bounds} holds for all $t \geq 0$. Furthermore, if the detectability and stabilizability conditions hold, then similar bounds hold for the steady-state solutions, i.e.,
\begin{align*}
P_{1} \preceq P_{2} \preceq P_{3}.
\end{align*}
\qed

\subsection{Proof of Theorem~\ref{theorem:steady-state_bounds}}

Theorem~\ref{theorem:Ahlswede_Winter} is employed to bound the sum of $n_s$ i.i.d. random matrices $Z$ outlined in Section \ref{section:problem_formulation},
\begin{align*}
\mathbb{P} \Bigg{[} (1-\epsilon)\mathbb{E}[Z] \preceq \frac{1}{n_s}\sum_{j=1}^{n_s} Z_{j} \preceq (1+\epsilon) \mathbb{E}[Z] \Bigg{]} &\geq (1-\delta),
\end{align*}
such that $\delta, \epsilon \in (0,1)$, $\delta = 2m \, {\mathrm{e}}^\frac{ -\epsilon^2 n_s }{ 4 \rho }$, and $\mathcal{Z}_{j} \preceq \rho \, \mathbb{E}[ Z ]$ for all $j \in \{ 1 , \ldots , n_c \}$. The event guaranteed with at least probability $(1 - \delta)$ is simplified below.
\begin{align}
\label{eqn:inequality_X}
(1-\epsilon) n_s \mathbb{E}[Z] \preceq C_{\mathcal{S}}^{T} R_{\mathcal{S}}^{-1} C_{\mathcal{S}}^{} \preceq (1+\epsilon) n_s \mathbb{E}[Z]
\end{align}
If $(1-\epsilon) n_s \mathbb{E}[Z]$, $C_{\mathcal{S}}^{T} R_{\mathcal{S}}^{-1} C_{\mathcal{S}}^{}$, and $(1+\epsilon) n_s \mathbb{E}[Z]$ are denoted as $Y_1$, $Y_2$, and $Y_3$, respectively, then, under the context of Lemma~\ref{lemma:arbitrary_model_Y1_Y2_Y3}, the filtered covariance matrix $P_{\mathcal{S},t-1}$ corresponding to $C_{\mathcal{S}}^{T} R_{\mathcal{S}}^{-1} C_{\mathcal{S}}^{}$ is upper and lower bounded, i.e.,
\begin{align}
\label{eqn:inequality_P}
P_{L,t-1} \preceq P_{\mathcal{S},t-1} \preceq P_{U,t-1}, \ \forall t \geq 0.
\end{align}
Since inequality \eqref{eqn:inequality_P} is derived from \eqref{eqn:inequality_X}, then
\begin{align}
\label{eqn:probability_P_t-1}
\mathbb{P} [ P_{L,t-1} \preceq P_{\mathcal{S},t-1} \preceq P_{U,t-1} ] &\geq (1-\delta), \ \forall t \geq 0.
\end{align}
Below we outline the necessary and sufficient conditions of Lemma~\ref{lemma:arbitrary_model_Y1_Y2_Y3} under the context of Thoerem~\ref{theorem:steady-state_bounds}. For instance, $P_{U,t-1}$ and $P_{L,t-1}$ converge to their respective steady-state solution if $(A, \mathbb{E}[Z]^{1/2})$ and $(A, Q^{1/2})$ are detectable and stabilizable, respectively. Similarly, $P_{\mathcal{S},t-1}$ converges to its steady-state solution if $(A,C_{\mathcal{S}})$ and $(A, Q^{1/2})$ are detectable and stabilizable, respectively. Furthermore, the initial estimate of filtered covariance matrices $P_{L,t-1}$, $P_{\mathcal{S},t-1}$, and $P_{U,t-1}$ must satisfy the inequality, $P_{1,-1} \preceq P_{2,-1} \preceq P_{3,-1}$, such that $P_{1,-1} \succeq 0$. If the filtered covariance matrices converge to a steady-state, then concentration inequality \eqref{eqn:probability_P_t-1} reduces to the following,
\begin{align}
\label{eqn:probability_P}
\mathbb{P} [ P_{L} \preceq P_{\mathcal{S}} \preceq P_{U} ] &\geq (1-\delta),
\end{align}
where $P_{U}$, $P_{L}$, and $P_{\mathcal{S}}$ denote the steady-state solution to $P_{U,t-1}$, $P_{L,t-1}$, and $P_{\mathcal{S},t-1}$, respectively, i.e.,
\begin{align*}
P_{U}^{-1} &= (A P_{U}^{} A^{T} + Q)^{-1} + (1-\epsilon) n_s \mathbb{E}[Z],   \\
P_{L}^{-1} &= (A P_{L}^{} A^{T} + Q)^{-1} + (1+\epsilon) n_s \mathbb{E}[Z],   \\
P_{\mathcal{S}}^{-1} &= (A P_{\mathcal{S}}^{} A^{T} + Q)^{-1} + C_{\mathcal{S}}^{T} R_{\mathcal{S}}^{-1} C_{\mathcal{S}}^{}.
\end{align*}
\qed

\subsection{Proof of Lemma~\ref{lemma:lower_bound_L_&_E_P}}
We define the steady-state error covariance $P_{\mathcal{S}}$ as the solution to the following,
\begin{align*}
P_{\mathcal{S}}^{-1} &= ( A P_{\mathcal{S}}^{} A^{T} + Q )^{-1} + C_{\mathcal{S}}^{T} R_{\mathcal{S}}^{-1} C_{\mathcal{S}}^{},
\end{align*}
and the expectation of $P_{\mathcal{S}}^{-1}$ as
\begin{align}
\label{eqn:E_P_S_-1}
\mathbb{E}[ P_{\mathcal{S}}^{-1} ] &= \mathbb{E}[ ( A P_{\mathcal{S}}^{} A^{T} + Q )^{-1} ] + n_s \mathbb{E}[Z].
\end{align}
We first derive an upper bound on $\mathbb{E}[ ( A P_{\mathcal{S}}^{} A^{T} + Q )^{-1} ]$.
\begin{align}
\label{eqn:upper_bound_E}
& \mathbb{E}[ ( A P_{\mathcal{S}}^{} A^{T} + Q )^{-1} ] \nonumber   \\
&\overset{(a)}{=} \mathbb{E}[ Q^{-1} -  Q^{-1} A ( P_{\mathcal{S}}^{-1} + A^{T} Q^{-1} A )^{-1} A^{T} Q^{-1} ] \nonumber   \\
&= Q^{-1} -  Q^{-1} A \mathbb{E}[ ( P_{\mathcal{S}}^{-1} + A^{T} Q^{-1} A )^{-1} ] A^{T} Q^{-1} \nonumber   \\
&\overset{(b)}{\preceq} Q^{-1} -  Q^{-1} A  ( \mathbb{E}[ P_{\mathcal{S}}^{-1} ] + A^{T} Q^{-1} A )^{-1} A^{T} Q^{-1} \nonumber   \\
&\overset{(c)}{\preceq} ( A (\mathbb{E}[ P_{\mathcal{S}}^{-1} ] )^{-1} A^{T} + Q )^{-1}
\end{align}
Convexity of function $f_{1}( P_{\mathcal{S}}^{-1} ) = ( P_{\mathcal{S}}^{-1} + A^{T} Q^{-1} A )^{-1}$ implies $f_{1}( \mathbb{E}[ P_{\mathcal{S}}^{-1} ] ) \preceq \mathbb{E}[ f_{1}( P_{\mathcal{S}}^{-1} ) ]$ over the cone of positive semi-definite matrices, according to Jensen's inequality \cite{jensen1906fonctions}. Step $(a)$ and $(c)$ hold by the matrix inversion lemma \cite{henderson1981deriving}, and $(b)$ holds by the matrix convexity of $f_{1}( P_{\mathcal{S}}^{-1} )$. Given equality \eqref{eqn:E_P_S_-1} and inequality \eqref{eqn:upper_bound_E}, we obtain the following bound on $\mathbb{E}[ P_{\mathcal{S}}^{-1} ]$,
\begin{align}
\label{eqn:E_P_t_-1}
\mathbb{E}[ P_{\mathcal{S}}^{-1} ] &\preceq ( A (\mathbb{E}[ P_{\mathcal{S}}^{-1} ] )^{-1} A^{T} + Q )^{-1} + n_s \mathbb{E}[Z].
\end{align}
Inequality \eqref{eqn:E_P_t_-1} suggests an upper bound on $\mathbb{E}[ P_{\mathcal{S}}^{-1}] $, denoted as $\mathbb{E}[ P_{\mathcal{S}}^{-1}]_{U} $, satisfying
\begin{align}
\label{eqn:E_P_U}
\mathbb{E}[ P_{\mathcal{S}}^{-1} ]_{U} 
&= ( A (\mathbb{E}[ P_{\mathcal{S}}^{-1} ]_{U})^{-1} A^{T} + Q )^{-1} + n_s \mathbb{E}[Z].
\end{align}
Convexity of inverse function $f_{2}( P_{\mathcal{S}}^{-1} ) = ( P_{\mathcal{S}}^{-1} )^{-1}$ implies $\mathbb{E}[ P_{\mathcal{S}}^{-1} ]^{-1} \preceq \mathbb{E}[( P_{\mathcal{S}}^{-1})^{-1} ] = \mathbb{E}[ P_{\mathcal{S}} ]$ over the cone of positive semi-definite matrices, according to Jensen's inequality \cite{jensen1906fonctions}. Since $f_{2}( P_{\mathcal{S}}^{-1} )$ is convex and $\mathbb{E}[ P_{\mathcal{S}}^{-1} ] \preceq \mathbb{E}[ P_{\mathcal{S}}^{-1} ]_{U}$, then $\mathbb{E}[ P_{\mathcal{S}}^{-1} ]_{U}^{-1} \preceq \mathbb{E}[ P_{\mathcal{S}} ]$. Lemma~\ref{lemma:lower_bound_L_&_E_P} denotes $\mathbb{E}[ P_{\mathcal{S}}^{-1} ]_{U}^{-1}$ as $L$ for clarity in notation. If $(A,\mathbb{E}[Z]^{1/2})$ and $(A,Q^{1/2})$ are detectable and stabilizable, respectively, then the following recursive equation,
\begin{align*}
L_{t}^{-1} = (A L_{t-1}^{} A^{T} + Q)^{-1} + n_s \mathbb{E}[Z],
\end{align*}
converges to $L$ as time instant $t$ tends to infinity.
\qed

\subsection{Proof of Theorem~\ref{theorem:opt_sampling_dist}}

Minimizing $\overline{\lambda}( P_{U} )$ is equivalent to maximizing $\underline{\lambda}( P_{U}^{-1} )$. Below outlines the problem of maximizing $\underline{\lambda}( P_{U}^{-1} )$ for a selected $\rho \in \big{[} 1 , \frac{n_s \epsilon^2}{4 \log{ ( 2m / \delta ) }} \big{)}$ and an arbitrarily small $\eta > 0$.
\begin{align}
\max_{ P_{U}^{-1} , \{ p_{i} \}_{i=1}^{n_c} } & \underline{\lambda}( P_{U}^{-1} ) \nonumber   \\
\textrm{s.t.} \hspace{5.75mm}
& P_{U}^{-1} \succeq \eta I_m \nonumber   \\
& \epsilon = \sqrt{ \frac{ 4 \rho }{ n_s } \log{ \frac{2m}{\delta} } } \in (0,1) \nonumber   \\
& \sum_{i=1}^{n_c} p_{i} = 1, \ p_{i} \geq 0 , \ \forall i \in \{ 1 , \ldots , n_c \} \nonumber   \\
& \mathbb{E}[Z] = \sum_{j=1}^{n_c} p_{j} \mathcal{Z}_{j} \nonumber   \\
& \mathcal{Z}_{j} \preceq \rho \, \mathbb{E}[Z] , \ \forall j \in \{ 1 , \ldots , n_c \} \nonumber   \\
& P_{U}^{-1} = (A P_{U}^{} A^{T} + Q)^{-1} + (1-\epsilon) n_s \mathbb{E}[Z] \nonumber
\end{align}
The conditions outlined in Theorem~\ref{theorem:steady-state_bounds} are satisfied by the above constraints. Inequality $P_{U}^{-1} \succeq \eta I_m$ guarantees a lower bound on the minimum eigenvalue of $P_U^{-1}$, denoted as $\eta$. Thus, $\eta^{-1}$ is an upper bound on the maximum eigenvalue of $P_U$. The last equality constraint in our optimization problem can be restated, per the matrix inversion lemma \cite{henderson1981deriving}, as the following,
\begin{align}
0
&= Q^{-1} + (1-\epsilon) n_s \mathbb{E}[Z] - P_{U}^{-1} \nonumber   \\
\label{eqn:constraint_0_P_U_-1}
& \hspace{4mm} - Q^{-1} A ( P_{U}^{-1} + A^{T} Q^{-1} A )^{-1} A^{T} Q^{-1}
\end{align}
Disregarding all constraints, the general eigenvalue problem of maximizing $\underline{\lambda}( P_{U}^{-1} )$ is equivalent to the SDP below.
\begin{align}
\max_{ \lambda }
\quad & \lambda \nonumber   \\
\textrm{s.t.} \hspace{1mm} \quad 
\label{eqn:constraint_P_U_-1_lambda}
& ( P_{U}^{-1} - \lambda I_m ) \succeq 0
\end{align}
Condition \eqref{eqn:constraint_0_P_U_-1} and \eqref{eqn:constraint_P_U_-1_lambda} reduce to the following,
\begin{align}
0 &\preceq Q^{-1} + (1-\epsilon) n_s \mathbb{E}[Z] - \lambda I_m \nonumber   \\
\label{eqn:P_U_-1}
& \hspace{4mm} - Q^{-1} A ( P_{U}^{-1} + A^{T} Q^{-1} A )^{-1} A^{T} Q^{-1}.
\end{align}
Inequality \eqref{eqn:P_U_-1} is formulated into a linear matrix inequality (LMI) constraint by the Schur complement method \cite{zhang2006schur},
\begin{align*}
\begin{bmatrix}
P_{U}^{-1} + A^{T} Q^{-1} A & ( A^{T} Q^{-1} )   \\
( A^{T} Q^{-1} )^{T} & Q^{-1} + (1-\epsilon) n_s \mathbb{E}[Z] - \lambda I_m
\end{bmatrix}
\succeq 0,
\end{align*}
such that $P_{U}^{-1} + A^{T} Q^{-1} A \succeq 0$. The latter inequality is a redundant constraint, since inequality $P_{U}^{-1} \succeq \eta I_m$ implies $P_{U}^{-1}~\succ~0$ and Assumption \ref{assumption:covariance_Q,R} in tandem with the Conjugation Rule \cite{tropp2015introduction} implies $A^{T} Q^{-1} A \succeq 0$. Maximizing $\underline{\lambda}( P_{U}^{-1} )$ can be formulated into an SDP, such that the constraints depend affinely on the decision variables.
\begin{align}
\max_{ \lambda , P_{U}^{-1} , \{ p_{i} \}_{i=1}^{n_c} }
\quad & \lambda \nonumber   \\
\textrm{s.t.} \hspace{7mm} \quad
& \lambda > 0 , \ P_{U}^{-1} \succeq \eta I_m , \ \{ p_{i} \}_{i=1}^{n_c} \in \Delta^{n_c} \nonumber   \\
& \epsilon = \sqrt{ \frac{ 4 \rho }{ n_s } \log{ \frac{2m}{\delta} } } \in (0,1) \nonumber   \\
&
\begin{bmatrix}
\Gamma_{1} & \Gamma_{2}   \\
\Gamma_{2}^{T} & \Gamma_{3}
\end{bmatrix}
\succeq 0 \nonumber   \\
& \Gamma_{1} = ( P_{U}^{-1} + A^{T} Q^{-1} A ) \nonumber   \\
& \Gamma_{2} = ( A^{T} Q^{-1} ) \nonumber   \\
& \Gamma_{3} = ( Q^{-1} + (1-\epsilon) n_s \mathbb{E}[Z] - \lambda I_m ) \nonumber   \\
& \mathbb{E}[Z] = \sum_{j=1}^{n_c} p_{j} \mathcal{Z}_{j} \nonumber   \\
& \mathcal{Z}_{j} \preceq \rho \, \mathbb{E}[Z] , \ \forall j \in \{ 1 , \ldots , n_c \} \nonumber
\end{align}
Let $( \lambda^{*} , P_{U}^{-1*} , \{ p_{i}^{*} \}_{i=1}^{n_c} )$ denote the tuple of decision variables that optimally maximizes $\underline{\lambda}( P_{U}^{-1} )$ for a selected $\rho$. Note that $\lambda^{*}$ is the globally maximum $\underline{\lambda}( P_{U}^{-1} )$ and $\lambda^{*-1}$ is the globally minimum $\overline{\lambda}( P_{U} )$. Theorem~\ref{theorem:opt_sampling_dist} states a concise, but equivalent, SDP formulation and denotes $P_{U}^{-1}$ as $X$ for clarity in notation.
\qed

\end{document}